\documentclass[smallextended]{svjour3}
\smartqed  
\usepackage{graphicx}
\usepackage{epsfig}
\usepackage{amsmath, amssymb}
\usepackage{fancyhdr}
\usepackage[normalem]{ulem}
\usepackage{color}
\usepackage{hanging}
\usepackage{hyperref}
\hypersetup{
  colorlinks=true,
}
\usepackage{url}
\usepackage{float, placeins}
\usepackage{mwe}
\usepackage[caption=false]{subfig}
\usepackage{morefloats}
\usepackage{empheq}
\usepackage{bm}
\usepackage{booktabs} 
\usepackage{multirow}
\usepackage{makecell}
\usepackage{cases}
\usepackage{tabularx}
\usepackage[top=1.1in,left=1.1in,bottom=1.1in,right=1.1in]{geometry}

\newcolumntype{Y}{>{\centering\arraybackslash}X}

\begin{document}
\hypersetup{linkcolor= blue , citecolor= magenta, urlcolor = magenta}

\title{Finely tuned models sacrifice explanatory depth}

\author{Feraz Azhar \and Abraham Loeb}

\institute{Feraz Azhar \at
              Department of Philosophy, University of Notre Dame, Notre Dame, IN 46556, USA and \\Black Hole Initiative, Harvard University, Cambridge, MA 02138, USA \\
              \email{fazhar@nd.edu}           
                         \and
           Abraham Loeb \at
              Black Hole Initiative, Harvard University, Cambridge, MA 02138, USA\\
              \email{aloeb@cfa.harvard.edu}
}

\maketitle
\vspace{-2cm}
\noindent(Dated: December 17, 2020)
\vspace{0.5cm}
\begin{abstract}
It is commonly argued that an undesirable feature of a theoretical or phenomenological model is that salient observables are sensitive to values of parameters in the model. But in what sense is it undesirable to have such `fine-tuning' of observables (and hence of the underlying model)? In this paper, we argue that the fine-tuning can be interpreted as a shortcoming of the explanatory capacity of the model: in particular it signals a lack of {\it explanatory depth}. In support of this argument, we develop a schema---for (a certain class of) models that arise broadly in physical settings---that quantitatively relates fine-tuning of observables to a lack of depth of explanations based on these models. We apply our schema in two different settings in which, within each setting, we compare the depth of two competing explanations. The first setting involves explanations for the Euclidean nature of spatial slices of the universe today: in particular, we compare an explanation provided by the big-bang model of the early 1970s (where no inflationary period is included) with an explanation provided by a general model of cosmic inflation. The second setting has a more phenomenological character, where the goal is to infer from a limited sequence of data points, using maximum entropy techniques, the underlying probability distribution from which these data are drawn. In both of these settings we find that our analysis favors the model that intuitively provides the deeper explanation of the observable(s) of interest. We thus provide an account that relates two `theoretical virtues' of models used broadly in physical settings---namely, a lack of fine-tuning and explanatory depth---and argue that {\it finely tuned models sacrifice explanatory depth}. 
\end{abstract}

\section{Introduction \label{SEC:Introduction}}

One hope, in light of what is commonly claimed to be the fine-tuning of our existence---as described by our current favored physical theories---is that there is some alternate theory that renders our existence less-finely tuned. Such a hope is not restricted to concerns about life however, for it arises for scientific theories more broadly: in particular, it arises in cases where salient observables are deemed to be sensitive to values of parameters that are not specified by the theory. If one interprets part of the role that theories play as that of providing an {\it explanation} for measured values of observables, then a natural question arises: if measured values of observables are finely tuned, does this signal a problematic feature of the corresponding explanation of these measured values? In this paper we generally answer this question in the affirmative and argue that the feature of such an explanation that is compromised by finely tuned observables is that of its {\it depth}.

There are two types of theories in which we will be interested, that reflect a broad range of physical settings. The first type encompasses dynamical theories. Such theories can be characterized by (i) equations of motion for dynamical variables and (ii) free parameters, whose values are not fixed by the theory, but which need to be fixed in order to yield agreement with observations. Such free parameters typically come in two varieties: first, as initial conditions for the dynamical variables mentioned in (i) and secondly, as `constants', which do not explicitly refer to the dynamical variables. In principle, the sensitivity of observables described by such a theory can be due to any combination of (i) and (ii). Our focus will be on sensitivity that arises due to free parameters as specified in (ii). Our formalism also applies to a second type of  scenario encountered in the physical sciences. In particular, it applies to those scenarios in which established theoretical frameworks have yet to be developed but where there exist phenomenological characterizations of data. In such scenarios one can again identify salient observables, and our interest is in the sensitivity of these observables to values of parameters in such phenomenological models.\footnote{In what follows (and given what we have just described) we will not need to make much of the distinction between theories and models, and so for the sake of generality we shall refer primarily to ``models'' (though we will revert to using ``theories'' in places where it is more natural to do so, such as in Sec.~\ref{SEC:PertinentSettings}).} (See Secs.~\ref{SEC:Explanation} and~\ref{SEC:PertinentSettings} for a more complete description of the types of settings in which we are interested.)

Such models, we contend, play a crucial role in {\it explanations} for observables. Of course, what we mean by an explanation needs to be made precise: as is well known, the literature on scientific explanation is vast and controversial (see, for an overview,~\cite{kitcher+salmon_89},~\cite{skow_16},~\cite{woodward_19}, and references therein). The concept of the {\it depth} of an explanation---one that has practical appeal---has received relatively less attention. (See, for example,~\cite{hitchcock+woodward_03},~\cite{ylikoski+kuorikoski_10}, and~\cite{weslake_10}.) We commit, in this paper, to an account of explanation that has precursors in the literature (see, for example,~\cite{woodward+hitchcock_03}), but our account of depth is novel. We show, in a way that has been largely ignored in the literature, that the depth of an explanation can be related to the degree of fine-tuning of an observable. In short, we provide a schema that captures the following intuition: the greater the ranges of values of parameters [as described in (ii) above] that yield the same value (or similar values) of some salient observable, the less-finely tuned is that observable, and the deeper is the explanation of the value of the observable. In this way, we provide insight into a key aspect of model choice by relating two important `theoretical virtues': namely the virtue of a lack of fine-tuning of salient observables and the virtue of a deep explanation.\footnote{For some background on theoretical virtues in the sciences, see~\cite{kuhn_77} and~\cite{mcmullin_82}. For more recent work, see, for example,~\cite{keas_18} and~\cite{schindler_18}.}

Our plan for this paper is as follows. In Sec.~\ref{SEC:ConnectingFTwithE} we describe, from a conceptual point of view, our rationale for connecting fine-tuning with explanatory depth. In Sec.~\ref{SEC:Explanation}, we describe quantitative aspects of our schema. We first describe an account of explanation suited to a broad class of physical settings. We then introduce a quantitative notion of fine-tuning of observables (where these observables depend on some specified set of parameters), adapting work by Azhar and Loeb~\cite{azhar+loeb_18}, and describe how this notion can be related to a measure of the depth of an explanation.\footnote{The explanandum is, in effect, the values possessed by the observables. Our measure of depth takes into account {\it all} parameters that describe an observable---though the measure is most sensitive to parameters that render the observable `finely tuned'. Also, the inclusion of a larger number of parameters will generally decrease the depth of the corresponding explanation. In this way, a lack of depth is related to both a sensitive dependence of possessed values of an observable on (values of) parameters, as well as the total number of parameters.} In Sec.~\ref{SEC:Applications} we apply our schema to two different examples. In particular, in Sec.~\ref{SEC:Flatness}, we illustrate how the schema yields results that agree with the putative consensus about the relatively deeper explanation provided by a general model of cosmic inflation~\cite{guth_81}---for the Euclidean nature of spatial slices today---compared to the explanation provided by the big-bang model of the 1970s (namely, a cosmological model that traces the evolution of the universe back to a singularity without encountering an inflationary period). In Sec.~\ref{SEC:Models}, we turn our attention to more phenomenological characterizations of data. We analyze a basic problem in which one looks to infer an underlying probability distribution from a finite set of data. We show how our schema, applied in such a way as to discriminate between two models, favors the intuitively correct model. We conclude in Sec.~\ref{SEC:Discussion}, where we provide further context for our results and include a comparison with salient work in the literature. 

In short, we present work that conceptually relates a notion of explanatory depth with a lack of fine-tuning, in a way that applies to a broad range of physical settings. This is combined with a quantitative analysis in which we introduce a measure of depth that can be used to discriminate between models. Our schema yields results that accurately reflect the types of practical judgements that have been made in the past, as well as those that may be useful in other novel physical settings.

\section{Linking fine-tuning to explanatory depth}\label{SEC:ConnectingFTwithE}

Broadly considered, at the core of the concept of fine-tuning lies a sensitive dependence of `outputs' (of a certain type) on `inputs' (of a certain type). The concept has, perhaps unsurprisingly, been described in different ways in the scientific literature. For example, a non-finely tuned phenomenon is sometimes described as one that is `stable' or `robust' or `natural'. These various descriptions do not necessarily express the same idea---and distinctions have indeed been drawn between them---but they share a common general theme: that of an invariance or preservation of some phenomenon to changes of a certain type.\footnote{\label{fn:Naturalness}`Stability' is often characterized as the invariance of a phenomenon to perturbations in the model that describes that phenomenon. The `Einstein static universe' was famously shown by Eddington to be `unstable'~\cite{eddington_30}. The equilibrium condition in this model-universe specifies a particular value of the mass density of matter (in terms of the cosmological constant). A slight increase in the mass density of matter away this value leads to a runaway contraction of the universe, whereas a slight decrease leads to a runaway expansion. `Robustness' is closely related to this characterization of stability, but depending on the particular context, it can have distinct or added features. One such feature, that arises for biological systems, is a ``slow degradation of a system's functions after damage, rather than catastrophic failure''~\cite[p.~1663]{kitano_02}. `Naturalness' is a concept that is commonly employed in particle physics settings. It describes a ``prohibition of sensitive correlations between widely separated energy scales''~\cite[p.~82]{williams_15}. (See also~\cite{butterfield_19} and~\cite{williams_19} for recent foundational and philosophical accounts.) On the issue of how naturalness and fine-tuning (understood as akin to `stability') may come apart, see~\cite[Sec. 3]{williams_15}.}

For the types of settings we will explore in this paper, a broad characterization of the fine-tuning of some phenomenon, $\mathcal{F}$, described by some model, $\mathcal{M}$, can be stated as follows: if circumstances in $\mathcal{M}$ were a little different, $\mathcal{F}$ would change significantly~\cite{azhar+loeb_18}. Of course, we need to make precise what we mean by ``circumstances'', ``change significantly'', as well as the nature of the phenomena considered. In any specific physical setting, there is no agreed upon way to do this, thus assessments of fine-tuning seem to carry with them a degree of subjectivity. However, there is a class of models that are commonly considered in physics, for which (at least) some of the concepts that enter into this broad characterization of fine-tuning can be made more precise. In particular, we mean those models that can be described as dynamical systems. For such systems, the relevant phenomena are observables that we can measure in experiments. The ``circumstances'' refer to the conjunction of the following (as mentioned in Sec.~\ref{SEC:Introduction}): (i) equations of motion for dynamical variables; (ii) initial conditions for the variables; (iii) dimensionless free parameters that aren't specified by the model, but which need to be fixed in order to extract a crisp prediction from the model. Perhaps the least tractable aspect of this broad characterization is how to make precise the notion of ``a little different''. That is, the task arises---for which, we admit, no general consensus exists---of defining and justifying a measure the captures this notion. In this paper, we will tackle some of these questions in specific physical settings that include the type of dynamical setting described above, as well as settings that have a more phenomenological flavor. 

A finely tuned model can reasonably be defined as one that contains at least one finely tuned phenomenon, but precisely which phenomena should count as being relevant to such an assessment is an interesting conceptual question. (See~\cite{maudlin_talk} for a discussion of this issue.) For some model, there are a variety of phenomena, some more intuitively important than others, that require circumstances to turn out `just so'. In considering the fine-tuning of life, for example, {\it particular lives} are finely tuned (circumstances need to turn out in very particular ways for {\it you} to arise) but these phenomena aren't the ones used to diagnose the fine-tuning of life, as understood within the context of our current best physical models (such as the standard models of particle physics and cosmology). It is the supposed sensitivity of life itself---to changes in observed values of parameters in these models---that determines this diagnosis and, moreover, that is often deemed to be in need of explanation. Considering models more generally, is it perhaps less straightforward (setting aside, of course, the issue of precisely what we mean by `life') to identify relevant phenomena. A pragmatic approach will look to glean such phenomena from the theoretical and observational context. We will proceed under the assumption that one has, indeed, identified salient phenomena.

This leads to one of the main tasks of this paper. Namely, the following question arises: what problems are presented by salient phenomena that are finely tuned---and, thus, by finely tuned models? One response to this question looks to undermine a key premise, namely, it looks to deny that fine-tuning can be unambiguously assigned as a feature of phenomena and models. (See, for example,~\cite{mcgrew+al_01} and the response in~\cite{koperski_05}.) The charge is that fine-tuning cannot be described in `objective' terms and so one could reasonably call into question any putative case of fine-tuning. (Of course, a {\it lack} of fine-tuning could also, for a similar reason, be called into question.) Though we agree that establishing some objective account of fine-tuning remains a challenge, we will nonetheless argue that remaining agnostic about the possibility of fine-tuning is the wrong stance to take. One can and {\it should} look to assess degrees of fine-tuning (we return to this claim below): the question of what to then do with this assessment is a separate issue. 

Indeed, there are settings where fine-tuning seems to exist but it is not considered to be especially problematic. An exemplar of such a model is that of deterministic chaos. Sensitive dependence of the dynamics on initial conditions is the defining feature of such a model. But there are other settings---settings that are not chaotic---where finely-tuned models spark the intuition that there is a less-finely tuned model that could account for some salient phenomenon. (One may, of course, worry about such a possibility for chaos as well, but we do not have a suggestion as to what type of less-finely tuned models could be implemented instead and so we will not discuss this possibility any further.) For example, Bialek~\cite{bialek_12} states the following:
\begin{quote}
``Physics, especially theoretical physics, is the search for concise mathematical descriptions of Nature \dots The dirty laundry of this enterprise is that our mathematical descriptions of the world have parameters. \dots if the predictions of the model are too sensitive to the exact values of the parameters, there is something vaguely unsatisfying about our claim to have explained things. Such strong parameter-dependent explanations are often called finely tuned, and we have grown to be suspicious of fine tuning. Experience suggests that if parameters need to be set to precise (or somehow unnatural) values, then we are missing something in our mathematical description of Nature''~\cite[p.~247]{bialek_12}.
\end{quote}
This quote alludes to a contention that underlies our thinking (and underlies our claim, in the previous paragraph, that fine tuning {\it should} be assessed): namely, that finely tuned models are a spur to model development. There are a variety of examples in the history of science, we contend, where less-finely tuned models supplant more-finely tuned models. A well-known modern example is provided by cosmic inflation, which supplanted the big-bang model of the 1970s (without an inflationary phase): this example will be further developed in Sec.~\ref{SEC:Flatness}.\footnote{\label{fn:examples}Other examples that, we believe, are amenable to such an analysis include: (i) Ptolemy's geocentric model being supplanted by subsequent non-geocentric models (see, for some background, Weinberg~\cite{weinberg_15}, who describes Ptolemy's model as finely tuned) and (ii) the development of quantum chromodynamics, which provided a unified framework---via an account of interactions between quarks and gluons---to understand the `zoo' of hadrons that had been discovered by nuclear physicists from the mid-twentieth century onward.}

Part of the reason why one might not find a finely tuned explanation of some phenomenon to be unsatisfying, is that the model that describes the phenomenon does, indeed, account for the phenomenon. That is, there is nothing ostensibly empirically inadequate about the model. Rather, as is evident in the above quote, what seems to require attention is that there is something distinctly problematic about the {\it explanation} into which the model enters. In this paper, we will argue that the explanatory virtue that is being compromised by such a model can be identified as the depth of the explanation it provides. 

Explanatory depth, unlike explanation has received relatively little attention in the literature. Moreover---and more relevantly for our purposes---the connection between explanatory depth and sensitivity, in the sense described above, is lacking a careful treatment. Our paper looks to address this gap. Of course, one aspect of establishing a link between fine-tuning and explanatory depth is the question of how we define an explanation. The account that we develop in the following section is a special case of the account of explanation developed by Woodward and Hitchcock~\cite{woodward+hitchcock_03}, but our account of depth has a significantly different emphasis. In particular, we focus on certain types of `explanatory generalizations': namely, those that correspond to equations of motion  for dynamical systems or else those (of a certain kind) that are non-dynamical but phenomenological in character. We then focus on an aspect of depth, namely, a lack of sensitivity, that is not captured by their subsequent work on depth (which appears in~\cite{hitchcock+woodward_03}). (Further comparative details appear in Sec.~\ref{SEC:HWcf}---the reader interested in this contrast may wish to consult this section now.) In a broader treatment of explanatory ``power'', Ylikoski and Kuorikoski~\cite{ylikoski+kuorikoski_10} describe various dimensions of explanatory power in the context of a particular account of explanation (though their account applies, in their words, to a notion of depth as well). One of these dimensions is indeed a lack of sensitivity and our account will look to make precise how this sensitivity can manifest in certain physical settings. For now, we turn to our account of such settings, and how we understand explanation, fine-tuning, and explanatory depth in these settings.

\section{Quantitative aspects of our schema \label{SEC:Explanation}}

In this section, we will describe our schema linking fine-tuning and explanatory depth, which consists of three parts: an account of relevant physical settings and the manner in which explanation operates in such settings; an account of fine-tuning; and a measure for explanatory depth. We turn now to the first part.

\subsection{Explanation in a broad class of physical settings}\label{SEC:PhysicalSettings}

The physical settings in which we are interested involve the following features.
\begin{itemize}
\item[(i)] A set of parameters, $\bm{p}=\{p_1, p_2, \dots, p_n\}$, which collectively take values in some (finite) subset, $\mathcal{P}$, of $n$-dimensional Euclidean space, $\mathbb{R}^{n}$ ($p_{i} \in \mathbb{R}$ for each $i=1,2,\dots,n$). These parameters will form part of the explanans. 
\item[(ii)] A set of observables, $\vec{O}=\{O_1, O_2, \dots, O_m\}$, which collectively take values in $\mathbb{R}^{m}$ ($O_{j} \in \mathbb{R}$ for each $j=1,2,\dots,m$). Specific values for this set of observables will correspond to the explananda. The observables can be thought of as  maps acting on the parameters described in (i): 
\begin{align}\label{EQN:ObsMapping}
\vec{O}:\;& \mathcal{P} \to \mathbb{R}^{m}\nonumber\\
& \bm{p} \mapsto \vec{O}(\bm{p}) \equiv \left\{O_{1}(\bm{p}), O_{2}(\bm{p}), \dots, O_{m}(\bm{p})\right\}. 
\end{align} 

We remain open as to the precise manner in which such maps might be realized, but in accordance with the two types of models that we will primarily focus on in this paper (as mentioned in both of the previous sections), such a map might represent the output of a dynamical system, in which case the parameters would then naturally correspond to initial conditions for any dynamical variables, as well as constants that are left unspecified by the model that underlies the dynamical system. Alternatively, such a map could arise for a model that implements a non-dynamical and phenomenological relationship between parameters and observables. We will explore models of both types in Sec.~\ref{SEC:Applications}. 
\end{itemize}
An explanation, on our account, will then be an argument in which the parameters take certain values, say, $\bm{p}^{\prime}$, such that the mapping in Eq.~(\ref{EQN:ObsMapping}) yields some specific set of values for the observables, say, $\vec{O}_{M}$. We can represent this argument in the following summarized form (with ``$E$'' for ``Explanation''): 
\begin{equation}\label{EQN:DetExplanation}
{E}: \bm{p}^{\prime} \wedge [\vec{O}(\bm{p}^{\prime})=\vec{O}_{M}] \therefore \vec{O}_{M}.
\end{equation}
(There are, indeed, similarities between this account of an explanation and that developed in~\cite{woodward+hitchcock_03}---we will further elaborate on this connection in Sec.~\ref{SEC:HWcf}.) 

As argued by Azhar and Loeb~\cite{azhar+loeb_18}, we justify the use of a finite parameter space by noting that the types of physical models for which our schema applies typically have finite regimes of applicability.\footnote{As we will discuss in Sec.~\ref{SEC:PertinentSettings}, our schema is applicable to physical settings described more generally by effective field theories (and, in particular, dynamical systems derived from such theories). These theories come with an energy cutoff that delimits the regime of applicability of the theory.} In such cases, these regimes of applicability limit the ranges of parameters. If there are no such available constraints, then we must advert to our expectations about what would comprise a reasonable range---of course, there is a degree of subjectivity in this assessment, but one that, we contend, can still play a useful role.

\subsection{A global perspective on fine-tuning}

As we mentioned above, we link an aspect of the depth of an explanation to a measure for fine-tuning. Azhar and Loeb~\cite{azhar+loeb_18} recently introduced two perspectives on fine-tuning---a local perspective and a global perspective---with corresponding measures for fine-tuning in each case. The perspective on fine-tuning that is relevant for our discussion is the global perspective. We now rehearse and adapt that account for our purposes. 

We will construct a measure for fine-tuning that allows one to compute the degree of fine-tuning of the observables with respect to each parameter {\it independently} (we focus on the $i$th parameter in what follows). (Our measure of depth will then combine these independent measures into a single one.) The measure makes a simple intuition precise. In short, the measure for fine-tuning of the observables with respect to the $i$th parameter compares (i) the range of values of the $i$th parameter that does not lead to a significant change in the values of the observables with (ii) the total possible range of values that the parameter can take. The smaller the range in (i), relative to the range in (ii), the greater the degree of fine-tuning.

Consider then, at some point ${\bm p}^{\prime}$ in parameter space $\mathcal{P}$, a direction, denoted by the unit vector $\hat{{\bm v}}_{i}^{+}$, where this vector points in the (positive) $i$th direction in parameter space. We need to find the minimum length of the vector, ${\bm v}_{i}^{+}$, that points in this direction, such that a significant change in the observables, computed at the point ${\bm p}^{\prime}+{\bm v}_{i}^{+}$, occurs. One way in which we can quantify such a change is, for example, via an order-unity fractional change in the observables (as defined in~\cite{azhar+loeb_18}). To be precise, we need to find $|{\bm v}_{i}^{+}|$ such that
\begin{equation}
\frac{|\vec{O}({\bm p}^{\prime}+{\bm v}_{i}^{+})-\vec{O}({\bm p}^{\prime})|}{|\vec{O}({\bm p}^{\prime})|}\sim \mathcal{O}(1).
\end{equation}
Another way in which we can consider a change in the observables to be significant is when any one of the observables that comprise $\vec{O}$ lies outside bounds established by experimental considerations.\footnote{\label{FN:sigma}At first glance, this definition of a ``significant change'' may seem problematic, in that it will lead to scenarios in which the smaller the experimental bounds on some observable---namely, the more precisely we can pinpoint its value---the more finely tuned will be that observable. We wish to make two points about this issue. First, as indeed pointed out in~\cite[p.~211]{ylikoski+kuorikoski_10} there is a tradeoff between precision (of the explanandum) and sensitivity: ``This is simply because smaller causal deviations are needed to disrupt the dependency between the {\it explanans} and a fine-grained {\it explanandum} than coarser-grained ones''. Second, an important dimension of our account is the comparative role that our account of depth can play. In particular, comparing the depth of two different explanations (supported by two different underlying models), for some observable whose value is determined within certain experimental bounds, is an important pragmatic aspect of our account.} In our examples in Sec.~\ref{SEC:Applications}, we will indeed study such scenarios.\footnote{Note that in the case where ${|\vec{O}({\bm p}^{\prime})|}=0$ [so that $O_{i}({\bm p}^{\prime})=0$ for each $i$] one would need to modify the schema above. In particular, a ``significant change'' in the observables then arises when a shift to the point ${\bm p}^{\prime}+{\bm v}_{i}^{+}$ yields a value for $|\vec{O}({\bm p}^{\prime}+{\bm v}_{i}^{+})|$, which is a significant fraction of the total distance that one could travel in the resultant direction in $\vec{O}(\mathcal{P})$, namely, in the image of the map $\vec{O}$ [in Eq.~(\ref{EQN:ObsMapping})].}

Having chosen a means to characterize changes in observables as significant, one then repeats the construction for the negative $i$th direction. Thus we find the length of the vector, ${\bm v}_{i}^{-}$, that points in this direction that leads to a significant change in the observables. The sum of the lengths of these two vectors will then correspond to the total range along the $i$th parameter direction over which observables do not change by a significant amount. Denote the size of this range by $|{\bm v}_{i}^{\pm}|$, where
\begin{equation}
|{\bm v}_{i}^{\pm}| \equiv |{\bm v}_{i}^{+}|+|{\bm v}_{i}^{-}|.
\end{equation}
Furthermore, let $\Delta[{\bm p}^{\prime}, \hat{{\bm v}}_{i}^{+}]$ denote the size of the range of {\it allowed} parameter values (as encoded in one's definition of $\mathcal{P}$) in the positive $i$th direction starting at ${\bm p}^{\prime}$, with $\Delta[{\bm p}^{\prime}, \hat{{\bm v}}_{i}^{-}]$ denoting the corresponding size in the negative $i$th direction. The sum of these two sizes, which we will denote by $\Delta[{\bm p}^{\prime}, \hat{{\bm v}}_{i}^{\pm}]$, where
\begin{equation}
\Delta[{\bm p}^{\prime}, \hat{{\bm v}}_{i}^{\pm}]\equiv \Delta[{\bm p}^{\prime}, \hat{{\bm v}}_{i}^{+}]+\Delta[{\bm p}^{\prime}, \hat{{\bm v}}_{i}^{-}],
\end{equation}
is the size of the allowed range considering both directions. The degree of global fine-tuning of the observables with respect to the $i$th parameter is then given by
\begin{equation}\label{EQN:GFT}
\mathcal{G}_{i}(\vec{O}; {\bm p}^{\prime})\equiv \log_{10}\left(\frac{\Delta[{\bm p}^{\prime}, \hat{{\bm v}}_{i}^{\pm}]}{|{\bm v}_{i}^{\pm}|}\right).
\end{equation}
Following~\cite{azhar+loeb_18}, we invoke the convention that if in some direction in parameter space, one reaches the edge of parameter space and no significant change in the observables has occurred, then the length of the vector that is encoded in the above formalism is the full distance that one can indeed travel in that direction.

This measure of (global) fine-tuning [in Eq.~(\ref{EQN:GFT})] has the following features.
\begin{itemize}
\item[(a)] $\mathcal{G}_{i}(\vec{O}; {\bm p}^{\prime})\geq 0$ since, by construction, $|{\bm v}_{i}^{\pm}| \leq \Delta[{\bm p}^{\prime}, \hat{{\bm v}}_{i}^{\pm}]$.
\item[(b)] The minimum value of $\mathcal{G}_{i}(\vec{O}; {\bm p}^{\prime})$ (namely, zero), occurs when $|{\bm v}_{i}^{\pm}| = \Delta[{\bm p}^{\prime}, \hat{{\bm v}}_{i}^{\pm}]$; the maximum value is unbounded. The minimum value corresponds to what we will call the {\it order-0 case}---in which the observables exhibit no fine-tuning, with regard to the corresponding parameter. The level of fine-tuning of the observables increases as $\mathcal{G}_{i}(\vec{O}; {\bm p}^{\prime})$ increases. 
\item[(c)] The use of the logarithm is so that (for example) order unity differences in the value of $\mathcal{G}_{i}(\vec{O}; {\bm p}^{\prime})$ can be thought of as being `significantly' different. For example, a case in which $\mathcal{G}_{i}(\vec{O}; {\bm p}^{\prime})=2$ is an order of magnitude more finely tuned than a case in which $\mathcal{G}_{i}(\vec{O}; {\bm p}^{\prime})=1$.
\end{itemize}

Note that fine-tuning of a phenomenon is often cast in terms of a low {\it probability} for that phenomenon. Our account of fine-tuning does not ostensibly involve probabilistic structure. A benefit of this approach is that it is often difficult to justify probability distributions over parameters. It is perhaps unsurprising that {\it given} a probability distribution over parameter space and, say, a probabilistic `law' that connects parameters to observables, one can generalize the above measure of global fine-tuning. For the sake of completeness, we present such a generalization in Appendix~\ref{SEC:AppendixA}. Of course, ambiguities that are averted by avoiding the introduction of probabilistic structure are replaced by a need to justify our use of a Euclidean measure over parameter space (and the space in which observables take values). Our choice of such Euclidean structure is largely guided by pragmatic considerations---namely, such structure is simple (and simple to implement) and, we contend, features in the types of assessments that occur (at least implicitly) in the course of model development (see, for example, the displayed quote in Sec.~\ref{SEC:ConnectingFTwithE} and the ensuing discussion).

\subsection{Describing a measure of depth}\label{SEC:DepthMeas}

Our measure of the depth of the explanation $E$, which we denote by $D_{E}(\vec{O}; {\bm p}^{\prime})$, combines the degree of fine-tuning of the observables with respect to each parameter direction into a single number. This is done in such a way that, ceteris paribus, adding a single parameter cannot increase the depth of the explanation. The measure is given by
\begin{equation}\label{EQN:Depth}
D_{E}(\vec{O}; {\bm p}^{\prime})\equiv\frac{1}{\displaystyle \prod_{i=1}^{n} \left[1+\mathcal{G}_{i}(\vec{O}; {\bm p}^{\prime})\right]},
\end{equation}
and has the following features.
\begin{itemize}
\item[(i)] It ranges between 0 and 1, namely, $0 < D_{E}(\vec{O}; {\bm p}^{\prime})\leq 1$. The lower bound is approached as the degree of fine-tuning of the observables is large with respect to at least one of the parameters [that is, for example, $\mathcal{G}_{1}(\vec{O}; {\bm p}^{\prime})\to\infty \implies D_{E}(\vec{O}; {\bm p}^{\prime})\to 0$].  The upper bound is saturated when the order-0 case is satisfied for {\it each} parameter [namely, $\forall i,\;\mathcal{G}_{i}(\vec{O}; {\bm p}^{\prime})=0 \iff D_{E}(\vec{O}; {\bm p}^{\prime})=1$]. 
\item[(ii)] Within the range described in (i), $D_{E}(\vec{O}; {\bm p}^{\prime})$ is strictly monotonically decreasing with an increasing level of fine-tuning with respect to any parameter.  
\item[(iii)] As mentioned above, ceteris paribus, adding another parameter (indexed by $k$, say) to the explanation, {\it always} reduces the depth of the explanation, unless the observables satisfy the order-0 case with respect to that parameter [namely, unless $\mathcal{G}_{k}(\vec{O}; {\bm p}^{\prime})=0$]. So, ceteris paribus, adding a parameter with respect to which the observables are (effectively) insensitive does not change the depth of the resulting explanation.
\item[(iv)] The form of our measure of depth is general. If we have two different models, say, with a common set of observables $\vec{O}_c$  (where, say, $\vec{O}_c$ is a proper subset of the observables described by each model), one could use our measure to compare the depth of the explanation provided by each model for (measured values of) just those observables. One could also use our measure to compare, for a single model that describes some vector of observables $\vec{O}=\{O_1, O_2,\dots\}$, the depth of the explanation provided for different observables in that vector: as in, for example, a comparison of the depth of the explanation for (the measured value of) $O_1$ vs. the depth of the explanation for $O_2$.
\item[(v)] The precise functional form of the measure that preserves the features listed above [namely, (i)--(iv)] is not unique. A monotonically increasing function $f$, say, could be inserted in the denominator in Eq.~(\ref{EQN:Depth}) and the features described above would continue to hold (so that one can identify a {\it family} of measures of depth). So we could have defined, for example, 
\begin{equation}
D_{E}(\vec{O}; {\bm p}^{\prime})\equiv \left({ \prod_{i=1}^{n} f\left[1+\mathcal{G}_{i}(\vec{O}; {\bm p}^{\prime})\right]}\right)^{-1},
\end{equation}
and maintained the above features. Our choice of definition in Eq.~(\ref{EQN:Depth}) has been made for the sake of simplicity. 
\end{itemize}

This measure captures salient features of intuitive accounts of explanatory depth as they arise in the types of physical settings described in Sec.~\ref{SEC:PhysicalSettings}. We have avoided specifying desiderata of such a measure in such a way as to attempt to uniquely fix the measure---this, we contend, leads to a false sense of rigor. Our measure captures the practical, context-dependent sense in which physicists often make assessments about fine-tuning and about the depth of an explanation. This last claim will be further supported via the explication of two distinct examples, to which we now turn.

\section{Applications of our schema \label{SEC:Applications}}

Our examples correspond to two (very) different settings in which explanations arise in physical settings. The first setting corresponds to one in which there is a theoretical framework for describing the system of interest---from which a dynamical system can be used to extract observables. The example we will pursue in this context is from cosmology---where we will compare two separate explanations for the Euclidean nature of spatial slices of our universe today. The second setting corresponds to one in which such theoretical frameworks do not (yet) exist but where phenomenological characterizations of observables have been developed. The example we will pursue in this context compares two explanations that characterize a stream of data using maximum entropy techniques. For both examples, we show that the explanation that one might expect to furnish a deeper explanation is indeed the one favored by an application of our schema.

\subsection{The flatness problem in cosmology \label{SEC:Flatness}}

The universe today is consistent with being spatially flat, that is, the geometry of space can be described as a three-dimensional Euclidean space. There are two different explanations of this observation that we will probe in this section. The first arises in the context of the big-bang model (BBM) of cosmology as it had been developed by the early 1970s---this model traces the evolution of the universe back to a singularity without invoking a period of cosmic inflation. The second arises, indeed, for a general model of cosmic inflation, in which the very early universe undergoes accelerated expansion for a short period of time~\cite{guth_81}. A key motivation for the introduction of cosmic inflation was to overcome perceived explanatory shortcomings in the BBM: one of those being the finely tuned nature of the spatial flatness of the universe today. This comparison thus serves as a particularly appropriate setting in which to probe our account of explanatory depth.

In describing both of these models, we will assume, for the sake of simplicity that spatial sections of the universe are homogeneous and isotropic from the outset. Newton's gravitational constant will be denoted by $G$ and we will set the speed of light to be unity, namely, $c\equiv 1$. We begin by first describing how the geometry of space evolves according to the BBM (see also~\cite{brawer_96}), before connecting this description to the schema introduced in Sec.~\ref{SEC:Explanation}. We will then repeat this analysis for the case of cosmic inflation. 

In our discussion of the BBM, we will consider an idealized setting in which the universe undergoes two sequential phases, where in each phase there is, in effect, a single (fluid) source that drives (and is driven by) the expansion of the universe. The first phase will be a radiation-dominated phase and this will be followed by a matter-dominated phase (where ``matter'' refers to pressureless `dust'). The radiation-dominated phase will last from some initial time $t_i$ until $t_T$, which will denote the time of matter-radiation equality (an equality of mass/energy density). The matter-dominated phase will last from $t_T$ until today, denoted by $t_0$. 

The evolution of a homogeneous and isotropic space in any such phase can be described, in the context of general relativity, by Friedmann's equations. If, as is conventional, we let the physical (spatial) separation of points be determined by the scale factor, $a(t)$, then the first Friedmann equation describes how the evolution of the scale factor is related to the mass density of the universe, $\rho(t)$, as well as the curvature of spatial sections, represented by $k$ ($k=0$ corresponds to the situation where space is Euclidean). In particular let $H(t)\equiv\dot{a}(t)/a(t)$ denote the Hubble expansion rate, where overdots will denote derivatives with respect to time. Then the first Friedmann equation is given by 
\begin{equation}\label{EQN:FR1}
 H^2(t)=\frac{8\pi G}{3}\rho(t)-\frac{k}{a^{2}(t)}. 
\end{equation}

We denote the critical density by $\rho_{\textrm{crit}}(t)\equiv 3H^2(t)/(8\pi G)$ [namely, the density such that spatial sections, according to Eq.~(\ref{EQN:FR1}), are flat]. Then we can define the key variable that will be relevant in our discussion below, namely, the dimensionless density parameter $\Omega(t)$:
\begin{equation}\label{EQN:DensityVar}
\Omega(t)\equiv\frac{\rho(t)}{\rho_{\textrm{crit}}(t)}.
\end{equation}
Note that $\Omega(t)=1$ corresponds to flat spatial sections.\footnote{Note also that in later connecting these introductory remarks to our account of explanation, the value of the dimensionless density parameter at some initial time will correspond to the ``parameter'' in the corresponding explanation (what we have earlier called ${\bm p^{\prime}}$), whereas the value of the dimensionless density parameter today will correspond to the ``observable'' (what we have earlier called $\vec{O}$). In what follows---keeping with conventions in the cosmology literature---we will continue to refer to the function defined in Eq.~(\ref{EQN:DensityVar}) as a (dimensionless density) parameter.} A useful quantity for our analysis below will be the deviation of $\Omega(t)$ from unity (namely, the deviation of spatial sections from flatness). A straightforward calculation gives, from Eqs.~(\ref{EQN:FR1}) and~(\ref{EQN:DensityVar}), that the fractional deviation from unity, of the dimensionless density parameter, is given by
\begin{equation}\label{EQN:omegaFrac}
\frac{\Omega(t)-1}{\Omega(t)}=\frac{3 k}{8\pi G\rho(t)a^{2}(t)}.
\end{equation}

The other equation we will require follows from the law of conservation of stress-energy and describes how the mass density of some (fluid) source evolves with time. This equation is given by
\begin{equation}\label{EQN:Matter}
\dot{\rho}(t) + 3H(t)\left[\rho(t)+P(t)\right]=0,
\end{equation}
where $P(t)$ is the pressure of the fluid. In what follows we will assume that for any fluid source the pressure is related to the mass density by an equation of state in which $P(t)=w\rho(t)$, where $w=1/3$ for radiation and $w=0$ for dust. Assuming that we know, at some final time $t_f$, the values of $\rho(t_f)$ and $a(t_f)$, we can integrate Eq.~(\ref{EQN:Matter}) to show, for $t<t_f$, that
\begin{equation}\label{EQN:Rho}
\rho(t)=\rho(t_f)\left[\frac{a(t_f)}{a(t)}\right]^{3(1+w)}.
\end{equation}
We can now combine Eqs.~(\ref{EQN:omegaFrac}) and~(\ref{EQN:Rho}) to show that for $t<t_f$,
\begin{equation}\label{EQN:projectback}
\frac{\Omega(t)-1}{\Omega(t)} = \left[\frac{a(t)}{a(t_f)}\right]^{1+3w} \frac{\Omega(t_f)-1}{\Omega(t_f)}.
\end{equation}

We now apply the above machinery to an analysis of our idealized version of the BBM, beginning from the present time, namely, $t_0$, and working our way back in time. Invoking Eq.~(\ref{EQN:projectback}) for the matter-dominated phase we have 
\begin{equation}\label{EQN:projectbackMD}
\frac{\Omega(t_T)-1}{\Omega(t_T)} = \frac{a(t_T)}{a(t_0)} \frac{\Omega(t_0)-1}{\Omega(t_0)}.
\end{equation}
For the matter-dominated phase it is straightforward to show, combining Eqs.~(\ref{EQN:FR1}) and~(\ref{EQN:Rho}), that 
\begin{equation}\label{EQN:aMD}
\frac{a(t)}{a(t_0)}=\left(\frac{t}{t_0}\right)^{2/3},
\end{equation}
so that Eq.~(\ref{EQN:projectbackMD}) becomes
\begin{equation}\label{EQN:projectbackMD2}
\frac{\Omega(t_T)-1}{\Omega(t_T)} = \left(\frac{t_T}{t_0}\right)^{2/3} \frac{\Omega(t_0)-1}{\Omega(t_0)}.
\end{equation}
[Note that for the sake of simplicity, in deriving the evolution of the scale factor in Eq.~(\ref{EQN:aMD}) we have assumed that $k=0$---such an assumption will suffice for our purposes in this section.] We can now connect the left-hand side of Eq.~(\ref{EQN:projectbackMD2}) to the earliest time we will consider in the context of the BBM, namely, $t_i$, by an application of Eq.~(\ref{EQN:projectback}) to the radiation-dominated phase. We find
\begin{equation}\label{EQN:projectbackRD}
\frac{\Omega(t_i)-1}{\Omega(t_i)} = \left[\frac{a(t_i)}{a(t_T)}\right]^{2} \frac{\Omega(t_T)-1}{\Omega(t_T)}.
\end{equation}
For the radiation-dominated phase, one can show, from Eqs.~(\ref{EQN:FR1}) and~(\ref{EQN:Rho}) (again assuming  $k=0$), that
\begin{equation}\label{EQN:aRD}
\frac{a(t)}{a(t_T)}=\left(\frac{t}{t_T}\right)^{1/2},
\end{equation}
so that Eq.~(\ref{EQN:projectbackRD}) becomes
\begin{equation}\label{EQN:projectbackRD2}
\frac{\Omega(t_i)-1}{\Omega(t_i)} = \frac{t_i}{t_T}\frac{\Omega(t_T)-1}{\Omega(t_T)}.
\end{equation}
Combining Eqs.~(\ref{EQN:projectbackMD2}) and~(\ref{EQN:projectbackRD2}) we thus obtain an expression for the fractional deviation from unity, of the density parameter, at the initial time $t_i$, as it relates to such a deviation today:
\begin{equation}\label{EQN:projectbackall}
\frac{\Omega(t_i)-1}{\Omega(t_i)} = \frac{t_i}{t_T}\left(\frac{t_T}{t_0}\right)^{2/3} \frac{\Omega(t_0)-1}{\Omega(t_0)}.
\end{equation}

To determine $t_T$, namely, the time corresponding to matter-radiation equality, we apply Eq.~(\ref{EQN:Rho}) to each component separately, so that (using the subscripts ``R'' for radiation and ``M'' for matter)
\begin{align}
\rho_{\rm R}(t)&=\rho_{\rm R}(t_0)\left[\frac{a(t_0)}{a(t)}\right]^4,\\
\rho_{\rm M}(t)&=\rho_{\rm M}(t_0)\left[\frac{a(t_0)}{a(t)}\right]^3.
\end{align}
Dividing these equations we find that
\begin{equation}
\frac{\rho_{\rm R}(t)}{\rho_{\rm M}(t)}=\frac{\rho_{\rm R}(t_0)}{\rho_{\rm M}(t_0)}\frac{a(t_0)}{a(t)}.
\end{equation}
At $t_T$, the left-hand side of the above equation is unity (by definition of $t_T$), whereas $a(t)$ on the right-hand side can be treated as though we are in the matter-dominated phase [as for Eq.~(\ref{EQN:aMD})]. Thus we find that
\begin{equation}\label{EQN:TimeT}
t_T = \left(\frac{\Omega_{\textrm{R},0}}{\Omega_{\textrm{M},0}}\right)^{3/2} t_0,
\end{equation}
where we have used the convention that $\Omega_{\textrm{i},0}\equiv{\rho_{\textrm{i}}(t_0)}/{\rho_{\textrm{crit}}(t_0)}$. Each term on the right-hand side of Eq.~(\ref{EQN:TimeT}) can either be read off or estimated from recent cosmological data.\footnote{The computation of $\Omega_{\textrm{R},0}$ is a little subtle. To estimate this latter quantity from recent cosmological data we assume that the energy density in radiation today, $\rho_{\textrm{R},0}$, can be related to the temperature, $T_0$, of radiation today (in the cosmic microwave background). One finds that $\Omega_{\textrm{R},0}\propto T_{0}^{4}/H_{0}^{2}$.} We find, using the above approximations, that $t_T\approx 3.1 \times 10^{4}$ years.

Recent cosmological data allows us to find an upper bound for the size of the density parameter today, $\Omega(t_0)$. In particular, the {\it Planck} Collaboration's most recent results on cosmological parameters~\cite{planck_18_CP} imply, for our purposes, that  
\begin{equation}
1-\Omega(t_0) = 0.001\pm 0.002.
\end{equation}
This implies that the magnitude of the fractional deviation from unity of the density parameter today [namely, the right-most term on the right-hand side of Eq.~(\ref{EQN:projectbackall})] must be constrained in the following way so that the present-day value of the density variable agrees with cosmological observations:
\begin{equation}
\left\lvert\frac{\Omega(t_0)-1}{\Omega(t_0)}\right\rvert <\frac{0.003}{0.997}\equiv\left\lvert\frac{\Omega(t_0)-1}{\Omega(t_0)}\right\rvert_{\textrm{max}}.
\end{equation}
Thus from Eq.~(\ref{EQN:projectbackall}) we obtain a constraint on the {\it initial} density variable:
\begin{equation}\label{EQN:BoundO}
\left\lvert\frac{\Omega(t_i)-1}{\Omega(t_i)}\right\rvert < \frac{t_i}{t_T}\left(\frac{t_T}{t_0}\right)^{2/3} \left\lvert\frac{\Omega(t_0)-1}{\Omega(t_0)}\right\rvert_{\textrm{max}}\equiv\epsilon.
\end{equation}
Note that assuming the time $t_i$ refers to a time that is very early in the history of the universe, the upper bound, $\epsilon$, expressed in Eq.~(\ref{EQN:BoundO}) is small. One finds therefore that
\begin{equation}
\frac{1}{1+\epsilon}<\Omega (t_i)<\frac{1}{1-\epsilon}.
\end{equation}
Thus the size of the entire range of values of $\Omega(t_i)$ that yield agreement with cosmological observations, which we will denote by $\Delta\Omega_i$, is given by 
\begin{equation}
\Delta\Omega_i \equiv \frac{2\epsilon}{1-\epsilon^2}. 
\end{equation}

We can now explicitly connect this analysis with the formalism described in Sec.~\ref{SEC:Explanation}. In particular, the parameters involved in the BBM's explanation of the flatness of the universe today, namely, ${\bm p}^{\prime}\to \Omega(t_i)$ and $\vec{O}\to\Omega(t_0)$. (Of course, the mapping between parameters and observables that enters into this explanation has a dynamical character.) Denote the maximum possible value of $\Omega(t_i)$ by $\Omega_{\textrm{max}}$ (where we will later consider situations in which $\Omega_{\textrm{max}}\geq 10$), the degree of global fine-tuning of the observable is given by [adapting Eq.~(\ref{EQN:GFT})]
\begin{equation}
\mathcal{G}(\Omega(t_0); \Omega(t_i))=\log_{10}\left(\frac{\Omega_{\textrm{max}}}{\Delta\Omega_i}\right)=\log_{10}\left(\frac{\Omega_{\textrm{max}}}{2\epsilon/(1-\epsilon^2)}\right),
\end{equation}
so that the depth of the explanation provided by the BBM, which we will denote by ${D}_{\textrm{BBM}}$ is [adapting Eq.~(\ref{EQN:Depth})]
\begin{equation}\label{EQN:BBMdepth}
{D}_{\textrm{BBM}}(\Omega(t_0); \Omega(t_i)) = \left[1+\log_{10}\left(\frac{\Omega_{\textrm{max}}}{2\epsilon/(1-\epsilon^2)}\right)\right]^{-1}.
\end{equation}

For the case of cosmic inflation, we assume that there is a period of $N$ e-folds of inflation that occur just before $t_i$, beginning at the time $t_{\textrm{Inf}}$, so that $a(t_i)=e^{N}a(t_{\textrm{Inf}})$. Over this period of time, we will assume that the mass/energy density of the universe was constant. Thus, invoking Eq.~(\ref{EQN:omegaFrac}) with $\rho(t_{\textrm{Inf}})=\rho(t_i)$, we find that
\begin{align}
\frac{\Omega(t_{\textrm{Inf}})-1}{\Omega(t_{\textrm{Inf}})} = e^{2N} \frac{\Omega(t_i)-1}{\Omega(t_i)},
\end{align}
so that we have the following present-day observational bound on the density parameter at the beginning of inflation:
\begin{equation}\label{EQN:InfBound}
\left\lvert\frac{\Omega(t_{\textrm{Inf}})-1}{\Omega(t_{\textrm{Inf}})} \right\rvert= e^{2N} \left\lvert\frac{\Omega(t_i)-1}{\Omega(t_i)}\right\rvert < e^{2N} \epsilon,
\end{equation}
where the right-hand side of Eq.~(\ref{EQN:InfBound}) uses Eq.~(\ref{EQN:BoundO}). We see immediately that to obtain agreement with cosmological observations today, the fractional deviation from unity, of the density parameter at the beginning of inflation, can be significantly greater than the deviation at $t_i$. Indeed, more specifically, if there is {\it any} inflation ($N>0$), then the initial allowable fractional deviation from unity of the density parameter, is greater for inflation than in the case of the BBM. 

From Eq.~(\ref{EQN:InfBound}), we can bound admissible values for $\Omega(t_{\textrm{Inf}})$ depending on the size of $e^{2N}\epsilon$ relative to unity:
\begin{numcases}{\frac{1}{1+ e^{2N}\epsilon} < \Omega(t_{\textrm{Inf}}) <}
   \Omega_{\textrm{max}} & $e^{2N} \epsilon\geq 1$ \label{EQN:firstcase}
   \\
   \textrm{min}\left\{\frac{1}{1- e^{2N}\epsilon}, \Omega_{\textrm{max}}\right\} & $e^{2N} \epsilon< 1$. \label{EQN:secondcase}
\end{numcases}
When $e^{2N} \epsilon\geq 1$, namely, $N \geq (1/2)\ln (\epsilon^{-1})$, the total number of $e$-folds of inflation will be said to be ``large'' ($N\to N_{\textrm{L}}$). When $e^{2N} \epsilon < 1$, namely, $N < (1/2)\ln (\epsilon^{-1})$, the total number of $e$-folds of inflation will be said to be ``small'' ($N\to N_{\textrm{S}}$). A value of $N = (1/2)\ln (\epsilon^{-1})$ will correspond to the ``critical number'' of $e$-folds ($N\to N_{\rm critical}$). As we shall see shortly (in a particular case), for $N > N_{\rm critical}$, the depth of the explanation provided by inflation jumps to be close to unity (compared with $N < N_{\rm critical}$), thus effectively solving the flatness problem. 

We can now connect this inflationary analysis with the formalism described in Sec~\ref{SEC:Explanation}. The relevant parameter involved in the explanation, provided by cosmic inflation, for the flatness of the universe today, is ${\bm p}^{\prime}\to\Omega(t_{\textrm{Inf}})$ and the observable of interest is, as for the case of the BBM, $\vec{O}\to\Omega(t_0)$. (Again, the mapping between parameters and observables that enters into this explanation has a dynamical character.) The degree of global fine-tuning of the observable depends on whether the total number of $e$-folds is large or small (according to our convention in the previous paragraph). We find [adapting Eq.~(\ref{EQN:GFT})] that
\begin{numcases}{\mathcal{G}(\Omega(t_0);\Omega(t_{\textrm{Inf}}))=}
   \log_{10}\left(\frac{\Omega_{\textrm{max}}}{\Omega_{\textrm{max}}-\frac{1}{1+ e^{2N}\epsilon}}\right) & $e^{2N} \epsilon\geq 1$ \label{EQN:firstcaseG}
   \\
   \log_{10}\left(\frac{\Omega_{\textrm{max}}}{ \textrm{min}\left\{\frac{1}{1- e^{2N}\epsilon}, \Omega_{\textrm{max}}\right\}-\frac{1}{1+ e^{2N}\epsilon}}\right)  & $e^{2N} \epsilon< 1$. \label{EQN:secondcaseG}
\end{numcases}
The depth of the explanation provided by inflation, which we will denote by ${D}_{\textrm{INF}}$ is thus [adapting Eq.~(\ref{EQN:Depth})]
\begin{subnumcases}{{D}_{\textrm{INF}}(\Omega(t_0);\Omega(t_{\textrm{Inf}}))=}
   \left[1+\log_{10}\left(\frac{\Omega_{\textrm{max}}}{\Omega_{\textrm{max}}-\frac{1}{1+ e^{2N}\epsilon}}\right)\right]^{-1} & $e^{2N} \epsilon\geq 1$ \label{EQN:firstcaseD}
   \\
   \left[1+\log_{10}\left(\frac{\Omega_{\textrm{max}}}{ \textrm{min}\left\{\frac{1}{1- e^{2N}\epsilon}, \Omega_{\textrm{max}}\right\}-\frac{1}{1+ e^{2N}\epsilon}}\right)\right]^{-1}  & $e^{2N} \epsilon< 1$. \label{EQN:secondcaseD}
\end{subnumcases}

In computing numerical estimates of the depth of the explanation for the flatness of the universe today, as described by the BBM and by cosmic inflation, we need to first estimate $\epsilon$ as it appears in Eq.~(\ref{EQN:BoundO}). Using $t_0= 13.8$Gyr and a fiducial value of $t_i=10^{-34}$s (which is a rough estimate of the time at which inflation ends), we find that $\epsilon = \mathcal{O}(10^{-52})$. This yields $N_{\textrm{critical}}\approx 60$. 

\begin{figure}[!t]
    \centering
    \includegraphics[width=0.75\textwidth]{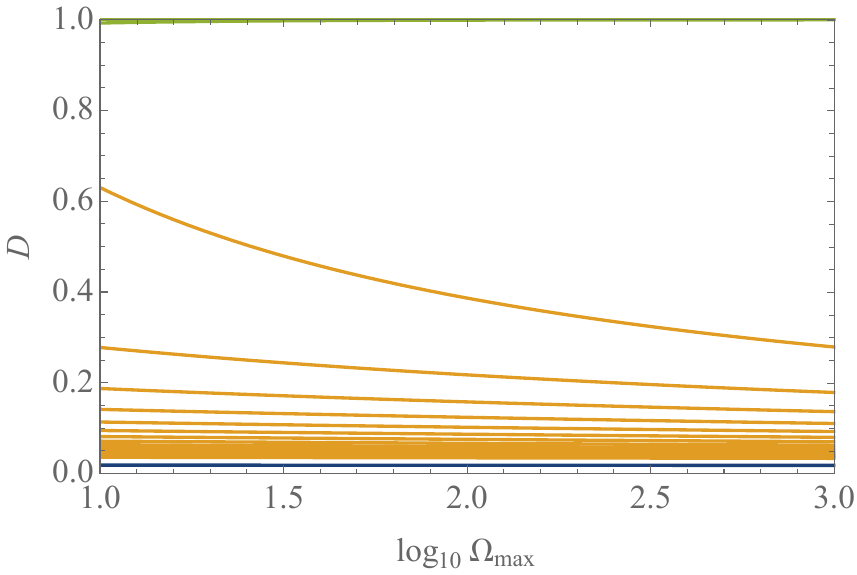}
    \caption{The depth, $D$, of the explanation for the flatness of spatial slices of our universe today as a function of the maximum possible value of the initial density parameter. The depth can vary, in principle, over the range $(0,1]$. The blue line corresponds to the depth as computed for the big-bang model (BBM) from Eq.~(\ref{EQN:BBMdepth}). The gold lines correspond to the depth as computed for different inflationary scenarios in which each successive gold line starting from the bottom of the figure corresponds to an increasing number of assumed $e$-folds of inflation, $N$, from $N=30$ through to $N=60$ (in steps of two $e$-folds). The green lines display the depth as computed for inflationary scenarios in which $N$ ranges from $N=61$ through to $N=91$ (in steps of 10 $e$-folds). We see that there is a distinct jump in the depth of the explanation when the number of $e$-folds is increased beyond a critical value (referred to as $N_{\textrm{critical}}$ in the main text). For the parameters chosen in the main text, $N_{\textrm{critical}}\approx 60$, in agreement with the generally accepted value for the total number of $e$-folds of inflation that is thought to solve various fine-tuning problems with the BBM (namely, problems that include the flatness problem). (See~\cite{azhar+butterfield_18} for further discussion of this point.)}
    \label{FIG:Depth}
\end{figure}

Figure~\ref{FIG:Depth} compares depths for the BBM and cosmic inflation as computed from Eqs.~(\ref{EQN:BBMdepth}),~(\ref{EQN:firstcaseD}), and~(\ref{EQN:secondcaseD}). We note that the depth of the BBM (in blue) is always (that is, for all chosen values of $\Omega_{\textrm{max}}$) less than the depth as computed in the case of inflation (gold and green lines). The gold lines represent the depth as it appears in Eq.~(\ref{EQN:secondcaseD}), namely, in cases where the number of $e$-folds of inflation is small (that is, less than the critical number of $e$-folds, $N_{\textrm{critical}}\approx 60$). The green lines represent the depth as it appears in Eq.~(\ref{EQN:firstcaseD}), namely, in cases where the number of $e$-folds of inflation is large (and greater than the critical number of $e$-folds).\footnote{ Note that increasing the number of $e$-folds of cosmic inflation well beyond the critical value ($N_{\rm critical}\approx 60$) does not (and indeed cannot) lead to a noticeable increase in the depth of the explanation (that is, the green lines in Fig.~\ref{FIG:Depth} effectively lie on top of each other near the maximum value of unity). This is consistent with the fact that the observable universe today circumscribes a horizon corresponding to the (last) 60 $e$-folds of inflation.}
Cosmic inflation thereby provides a deeper explanation for the flatness of our universe today compared with the BBM (as is commonly claimed).

\subsection{Model selection in maximum-entropy modeling \label{SEC:Models}}

We now turn to another example with a different character from the one described in Sec.~\ref{SEC:Flatness}. In particular, this example probes the depth of explanations provided by phenomenological models of data. The basic problem is as follows: one observes data presented as a sequence of $M$ samples, say, $x_1, x_2,\dots, x_M$, where we know that each sample is drawn independently from the same underlying probability distribution $P(x)$. The question of interest is: what is the probability distribution? In the absence of any further information we have no way of determining a unique distribution. But we can imagine constraining possibilities for such distributions by demanding that they satisfy constraints derived from the data. Of course, the set of such possible distributions is biased by the chosen constraints, but there is a class of distributions---known as maximum entropy distributions---that provide a useful (and relatively principled) guide.

Maximum entropy distributions, in short, are those that maximize the classical, Shannon-information theoretic entropy, subject to specified constraints. More formally, when the (random) variable of interest is defined over some continuous domain $\mathcal{D}$, the maximum entropy distribution is constructed by maximizing, over all functions $P(x) \geq 0$, the differential entropy $S[P(x)]$, where
\begin{equation}\label{EQN:DiffEntropy}
S[P(x)]\equiv -\int_{\mathcal{D}}dx\; P(x)\ln P(x),
\end{equation}
subject to $K$ constraints defined through some $K$ functions $f_{i}(x)$:
\begin{equation}\label{EQN:MEconstraints}
\int_{\mathcal{D}}dx\; P(x) f_{i}(x)=\mu_{i},
\end{equation}
for $i=1,2,\dots,K$. In addition, we must also enforce the usual normalization constraint:
\begin{equation}\label{EQN:MEnorm}
\int_{\mathcal{D}}dx\; P(x) = 1.
\end{equation}
A well-known theorem (see~\cite[Ch.~11]{cover+thomas_91}) provides the answer to this optimization problem:
\begin{equation}\label{EQN:ME}
P(x;\alpha_0, \alpha_1\dots,\alpha_K) = e^{\alpha_0 + \sum_{i=1}^{K} \alpha_{i} f_{i}(x)},
\end{equation}
where $\alpha_{0}, \alpha_{1}, \dots, \alpha_{K}$ need to be chosen so that Eqs.~(\ref{EQN:MEconstraints}) and~(\ref{EQN:MEnorm}) are satisfied. Note that in what follows, we use semicolons to delimit values of the underlying random variable (that is, $x$, above), from parameters on which the distribution depends (that is, $\alpha_0\dots,\alpha_K$, above).

From an interpretative point of view, these distributions [as represented by Eq.~(\ref{EQN:ME})] are the distributions that maximize our uncertainty (understood in information theoretic terms) about the underlying data given the constraints. In this sense they are the `least structured' distributions, consistent with the constraints. There has been renewed interest in their use, for example, in the biophysical sciences, especially in the case of underlying random variables that are discrete, and where the constraints of interest are of a particularly simple type.\footnote{We refer here to maximum entropy distributions that are consistent with one- and two-point correlation functions derived from the data (we will indeed probe such scenarios in the main text---though where the underlying random variable is continuous). There are a variety of biophysical systems that have been explored in this way, including: neural systems, proteins, the immune system, and even aggregations of birds (see~\cite{tkacik+bialek_16} for a review).}

We will describe a toy example that captures the sense in which our measure of explanatory depth can play a role in choosing between maximum entropy models. Consider the scenario in which we draw $M$ samples, $x_1, x_2,\dots, x_M$, from a known continuous probability distribution $P(x)$. The distribution we choose is the exponential distribution and takes the form
\begin{equation}\label{EQN:ExpD}
P(x) = \mu \exp\left({-\mu x}\right)\qquad x \geq 0, \mu>0.
\end{equation}
In what follows, to be concrete, we choose $\mu=1$. We will look to fit two different maximum entropy distributions to the data and will compare the depth of the explanations---about which we will say more shortly---provided by these distributions. The first distribution will be the maximum entropy distribution consistent with the empirical mean of the sample, denoted by $\langle x \rangle_{\rm e}$ (``e'' for ``empirical''). This distribution thus maximizes Eq.~(\ref{EQN:DiffEntropy}), for $\mathcal{D}\equiv [0,\infty)$, subject to a single constraint, as in Eq.~(\ref{EQN:MEconstraints}) with $f_1(x)\equiv x$ and $\mu_1\equiv\langle x \rangle_{\rm e}$. From Eq.~(\ref{EQN:ME}), this distribution takes a form that we will write as 
\begin{equation}\label{EQN:P1}
P_{1}(x;\lambda_0, \lambda_1) = \exp\left(-1-\lambda_0-\lambda_1 x\right). 
\end{equation}
Of course, this distribution has the {\it same functional form} as that from which the data are drawn [namely, Eq.~(\ref{EQN:ExpD})], and so we expect (given that there is a clear sense in which this is the `correct' distribution) the depth of the explanation of the data, derived from this distribution, to be relatively high.

The second distribution we will consider will be a maximum entropy distribution consistent with the first two empirically derived moments, namely, $\langle x \rangle_{\rm e}$ and $\langle x^2 \rangle_{\rm e}$. This distribution thus maximizes Eq.~(\ref{EQN:DiffEntropy}), again for $\mathcal{D}\equiv [0,\infty)$, subject to two constraints that take the form of Eq.~(\ref{EQN:MEconstraints}): one constraint has  $f_1(x)\equiv x$ and $\mu_1\equiv\langle x \rangle_{\rm e}$, whereas the second constraint has $f_2(x)\equiv x^2$ and $\mu_2\equiv\langle x^2 \rangle_{\rm e}$. From Eq.~(\ref{EQN:ME}), this maximum entropy distribution takes the form 
\begin{equation}\label{EQN:P2}
P_{2}(x;\lambda_0, \lambda_1, \lambda_2) = \exp\left(-1-\lambda_0-\lambda_1 x - \lambda_2 x^2\right). 
\end{equation}

In each case, the empirically derived moments will also have empirically derived standard deviations associated with them so that each moment that we consider is properly represented as 
\begin{equation}
\langle x^{k} \rangle_{\textrm{e}}\pm\sigma_k,\\
\end{equation}
where
\begin{equation}\label{EQN:SD}
\sigma_k \equiv \left[\frac{1}{M(M-1)}\sum_{i=1}^{M}\left(x_{i}^{\;k}-\langle x^k \rangle_{\textrm{e}}\right)^{2}\right]^{1/2},
\end{equation}
and where $x_{i}^{\;k}$ refers to the $i$th sample (out of the $M$ samples) raised to the $k$th power. 

In what sense, now, do maximum entropy distributions furnish an explanation of the data? The explanandum in which we will be interested will not correspond to the exact set of samples obtained but corresponds to the statistics of the sequence of samples as characterized by the {\it first two moments}. The rationale for this is that the precise set of samples are {\it representative} of the underlying distribution and what is salient about those samples is aggregate properties as captured by statistics that are deemed relevant. The first two moments represent two particularly simple features of the statistics of the data. 

In the case of the first maximum entropy model, represented by $P_{1}(x;\lambda_0, \lambda_1)$, the parameters involved in the explanation are given by ${\bm p}^{\prime} \to \left(\lambda_{0} , \lambda_1\right)$ and the observables are given by $\vec{O}\to\left(\langle x \rangle_{\rm e}, \langle x^2 \rangle_{\rm e}\right)$. The explanation is non-dynamical and phenomenological in character. One can represent the explanation, denoted by $E_1$, in the following way:
\begin{equation}\label{EQN:ME1}
E_1\;{\rm :}\;(\lambda_0, \lambda_1) \wedge\left[\int_{0}^{\infty}dx\;P_{1}(x;\lambda_0, \lambda_1) = 1 , \int_{0}^{\infty}dx\;P_{1}(x;\lambda_0, \lambda_1)x = \langle x \rangle_{\rm e}\right] \therefore (\langle x \rangle_{\rm e},  \langle x^2 \rangle_{\rm e}\pm\sigma_2).
\end{equation}
Note that in this first case, we relax the precision with which one needs to account for the second moment $\langle x^2 \rangle_{\rm e}$, as $P_{1}(x;\lambda_0, \lambda_1)$ is not defined in such a way as to demand agreement with the precise value of the second moment [unlike in the case of $P_2(x;\lambda_0, \lambda_1, \lambda_2)$]. Thus in what follows, we allow for the explananda in this first case to correspond, effectively, to $\vec{O}\to\vec{O}_1\equiv\left(\langle x \rangle_{\rm e}, \langle x^2 \rangle_{\rm e}\pm\sigma_2\right)$.

Similarly, in the case of the second maximum entropy model, represented by $P_{2}(x;\lambda_0, \lambda_1, \lambda_2)$, the parameters are given by ${\bm p}^{\prime} \to \left(\lambda_0, \lambda_1, \lambda_2 \right)$ and the observables are given by $\vec{O}\to\left(\langle x \rangle_{\rm e}, \langle x^2 \rangle_{\rm e}\right)$. One can represent the explanation, denoted by $E_2$, in the following way:
\begin{align}\label{EQN:ME2}
E_2\;{\rm :}\;(\lambda_0, \lambda_1, \lambda_2) \wedge&\left[
\int_{0}^{\infty}dx\;P_{2}(x;\lambda_0, \lambda_1,\lambda_2) = 1, 
\int_{0}^{\infty}dx\;P_{2}(x;\lambda_0, \lambda_1, \lambda_2)x = \langle x \rangle_{\rm e},\right. \nonumber\\
&\qquad\qquad\left.
 \int_{0}^{\infty}dx\;P_{2}(x;\lambda_0, \lambda_1, \lambda_2)x^2 = \langle x^2 \rangle_{\rm e} \right]\therefore (\langle x \rangle_{\rm e},  \langle x^2 \rangle_{\rm e}).
\end{align}
Our goal then is to compare the depth of the explanations provided by $E_1$ and $E_2$, which we will denote by $D_{E_1}$ and $D_{E_2}$, respectively.

Let us begin by analyzing $E_1$. In this case, one can solve for the parameter values $\lambda_0$ and $\lambda_1$ analytically. In particular, we'll rewrite the distribution in Eq.~(\ref{EQN:P1}) in the following way:
\begin{equation}\label{EQN:P1Z}
P_{1}(x;Z_1, \lambda_1)\equiv\frac{1}{Z_1}\exp\left(-\lambda_1 x\right),
\end{equation}
where $Z_1 \equiv \exp\left(1+\lambda_0\right)$. Enforcing the normalization constraint [namely, the first equation in square brackets in Eq.~(\ref{EQN:ME1})] gives $Z_1\to Z_{1:{\rm sol}}(\lambda_{1}) \equiv {1}/{\lambda_1}$; then enforcing the remaining constraint  [namely, the second equation in square brackets in Eq.~(\ref{EQN:ME1})] yields $\lambda_1\to \lambda_{1:{\rm sol}}\equiv 1/\langle x \rangle_{\rm e}$. Thus we find
\begin{equation}\label{EQN:P1Zsol}
P_{1}(x;Z_{1:{\rm sol}}(\lambda_{1:{\rm sol}}),\lambda_{1:{\rm sol}})=\frac{1}{\langle x \rangle_{\rm e}}\exp\left(-\frac{x}{\langle x \rangle_{\rm e}}\right).
\end{equation}
Now, we interpret this distribution as explaining the data if (and only if) it is also the case that the second moment of this distribution satisfies the corresponding experimental constraint, namely, that $\int_{0}^{\infty}dx\;P_{1}(x;Z_{1:{\rm sol}}(\lambda_{1:{\rm sol}}),\lambda_{1:{\rm sol}}) x^2 \in \langle x^2 \rangle_{\rm e}\pm\sigma_2$. Assuming this condition is satisfied (which it indeed will be in the numerical work that follows), we can compute the depth of the resulting explanation by computing the range of $\lambda_1$-values over which the empirical constraints continue to hold. We can determine this range through the following steps.
\begin{itemize}
\item[(i)] Fix $\delta$ to be some small positive real-valued number (small enough so that the following steps do not terminate at the first iteration in $k$). 
\item[(ii)] Fix $k=1$ and compute each of the following quantities:
\begin{align}
 \lambda_{1}^{*}&\equiv\lambda_{1:{\rm sol}}+k\delta,\nonumber\\
\langle x \rangle_{P_{1}^{*}}&\equiv\int_{0}^{\infty}dx\;P_{1}(x;Z_{1:{\rm sol}}(\lambda_{1}^{*}), \lambda_{1}^{*})\;x,\nonumber\\
\langle x^2 \rangle_{P_{1}^{*}}&\equiv\int_{0}^{\infty}dx\;P_{1}(x;Z_{1:{\rm sol}}(\lambda_{1}^{*}), \lambda_{1}^{*})\;x^2.\nonumber
\end{align}
\item[(iii)] If $\langle x \rangle_{P_{1}^{*}}\not\in \langle x \rangle_{\rm e}\pm\sigma_1$ and/or  $\langle x^2 \rangle_{P_{1}^{*}}\not\in \langle x^2 \rangle_{\rm e}\pm\sigma_2$ then the range of admissible values of $\lambda_{1}$ (greater than $\lambda_{1:{\rm sol}}$) is delineated by $\lambda_{1}^{*}$ as determined from the last value of $k$ for which the constraints were satisfied. \vspace{.1cm}

If the constraints are satisfied at this step, increment $k$ by 1 and repeat steps (ii) and (iii) until at least one of the constraints is indeed violated.
\item[(iv)]  Repeat steps (i) to (iii) starting at $k=-1$ and decrementing $k$ with each iteration to determine the range of admissible values of $\lambda_{1}$ less than  $\lambda_{1:{\rm sol}}$.
\end{itemize} 
This yields a range, $\delta\lambda_1$, around the value $\lambda_{1:{\rm sol}}$, over which the empirical constraints hold. The depth of the explanation is then given by
\begin{equation}\label{EQN:DE1}
D_{E_1}=\left[1+\log_{10}\left(\frac{1}{\delta\lambda_1}\right)\right]^{-1},
\end{equation}
where a fiducial value of unity for the possible range of $\lambda_1$ has been chosen to facilitate comparison. 

Let us now move on to the case of $E_2$. We rewrite the distribution in Eq.~(\ref{EQN:P2}) in the following way:
\begin{equation}\label{EQN:P2Z}
P_{2}(x;Z_2, \lambda_1, \lambda_2) = \frac{1}{Z_2}\exp\left(-\lambda_1 x - \lambda_2 x^2\right),
\end{equation}
where $Z_2\equiv\exp\left(1+\lambda_0\right)$. Enforcing the normalization constraint [namely, the first equation in square brackets in Eq.~(\ref{EQN:ME2})] allows us to consider $Z_{2}$ expressly as a function of $\lambda_1$ and $\lambda_2$, namely, $Z_{2}\to Z_{2:{\rm sol}}(\lambda_1, \lambda_2)$. In solving for the remaining free parameters in Eq.~(\ref{EQN:P2Z}) (namely, in computing $\lambda_{1:{\rm sol}}$ and $\lambda_{2:{\rm sol}}$) note the theorem by Dowson and Wragg~\cite[Theorem 3, p.~691]{dowson+wragg_73} that states, in our context, that if $\langle x^2 \rangle_{\rm e} > 2 \langle x \rangle_{\rm e}^2$, there is no distribution that maximizes the entropy. So in cases where $\langle x^2 \rangle_{\rm e} \leq 2 \langle x \rangle_{\rm e}^2$ we implement a simple numerical procedure to establish the desired maximum entropy distribution, which we denote by
\begin{equation}\label{EQN:P2Zsol}
P_{2}(x;Z_{2:{\rm sol}}(\lambda_{1:{\rm sol}}, \lambda_{2:{\rm sol}}), \lambda_{1:{\rm sol}}, \lambda_{2:{\rm sol}}) = \frac{1}{Z_{2:{\rm sol}}(\lambda_{1:{\rm sol}}, \lambda_{2:{\rm sol}})}\exp\left(-\lambda_{1:{\rm sol}} x - \lambda_{2:{\rm sol}} x^2\right).
\end{equation}
With such a distribution in hand, one can now compute the depth of the resulting explanation of the first two empirically derived moments by separately computing the range of $\lambda_1$-values over which the empirical constraints continue to hold (that is, while holding $\lambda_2$ fixed), followed by the range of $\lambda_2$-values over which the empirical constraints continue to hold (that is, holding $\lambda_1$ fixed). The procedure we follow mimics, in a straightforward way, the procedure we outlined for $E_1$, and yields ranges over which the constraints hold, namely, $\delta\lambda_1$ around the value $\lambda_{1:{\rm sol}}$, and $\delta\lambda_2$ around the value $\lambda_{2:{\rm sol}}$. The depth of the explanation is then given by
\begin{equation}\label{EQN:DE2}
D_{E_2}=\left[1+\log_{10}\left(\frac{1}{\delta\lambda_1}\right)\right]^{-1}\left[1+\log_{10}\left(\frac{1}{\delta\lambda_2}\right)\right]^{-1},
\end{equation}
 where, again, a fiducial value of unity for the possible range of $\lambda_{i}$ (with $i=1,2$) has been chosen to facilitate comparison.
 
The results of our trials are summarized in Table~\ref{TAB:MEtable}. We note that for any number of samples $M$, ${E_1}$ provides a deeper explanation than does ${E_2}$, and in many cases ${E_2}$ does not furnish an explanation at all. Thus the explanation that references the distribution from which the samples are actually drawn is comparatively better---and increasing the number of parameters by a single parameter (as for ${E_2}$) {\it decreases} the depth of the explanation (or, indeed, invalidates it entirely). 
\begin{table}
\begin{tabularx}{\linewidth}{@{}YYYYY@{}}
\hline\hline
&\multicolumn{2}{c}{{\it Random seed 1}} & \multicolumn{2}{c}{{\it Random seed 2}}\\
\cline{2-5}
{\it \small Number of samples} $M$  & $D_{E_1}$ & $D_{E_2}$ & $D_{E_1}$ & $D_{E_2}$ \\
\hline
$10^2$     & 0.578	&  	+    		&0.538	&0.312	\\
$10^3$     & 0.450	& 	+      		&0.439	&0.166	\\
$10^4$     & 0.362	& 	0.111    	&0.347	&0.111	\\
$10^5$ 	& 0.312	& 	+       	&0.312	&+	\\
$10^6$ 	& 0.251	& 	+       	&0.270	&+	\\
$10^7$ 	& 0.222	& 	+       	&0.228	&+	\\
\hline\hline
\end{tabularx}
\caption{\label{TAB:MEtable} Depth of two explanations ($D_{E_1}$ and $D_{E_2}$) in our maximum entropy toy example. We probe scenarios where the number of samples ranges over five orders of magnitude. For a fixed number of samples, we sample from the exponential distribution in Eq.~(\ref{EQN:ExpD}) (in one of two ways, using different random seeds to generate the sequence) and then look to explain statistics of the samples. In particular, we look to explain the first and second moments via one of two explanations, ${E_1}$ or ${E_2}$, as described in Eqs.~(\ref{EQN:ME1}) and~(\ref{EQN:ME2}), respectively. Results quoted as `+' refer to cases where a maximum entropy distribution (and hence an explanation) does not exist (see the discussion in the main text). In each case, we see that ${E_1}$ provides a deeper explanation than ${E_2}$, or else ${E_2}$ does not furnish an explanation at all. The decrease in the depth of the explanation as the number of samples increases arises because of the manner in which we have constrained errors in the statistics of samples [see Eq.~(\ref{EQN:SD})]: larger samples have smaller standard deviations, and thus a smaller target range for the observables (which the underlying parameters help to explain). (See fn.~\ref{FN:sigma} for a related discussion.)}
\end{table}

\section{Discussion\label{SEC:Discussion}}

In this paper, we have analyzed the intuition that an increased level of (global) fine-tuning of observables---and thus (by definition) of fine-tuning of the model that describes those observables---is associated with a deficiency in the explanation provided by the model. In particular, we claim that such fine-tuning signals a lack of depth of the explanation. The schema we have developed is general in that it can be applied to a broad range of scenarios across the physical sciences, where established theoretical frameworks exist and also in phenomenological settings where established theoretical frameworks do not yet exist. In this final section, we will (i) further describe aspects of the physical settings in which, we contend, our schema applies; (ii) comment on how our schema relates to work by Hitchcock and Woodward~\cite{hitchcock+woodward_03} on explanatory depth; and (iii) highlight further conceptual issues. 

\subsection{Pertinent physical settings}\label{SEC:PertinentSettings}

A natural question that arises is: what are the ``broad range of scenarios'' in which our schema applies? We contend that our schema is suited to the analysis of non-fundamental theories or, for example, {\it effective} theories (so, theories that aren't presented as `final theories'), where we are uncertain about the nature of the theory that applies at higher energy scales. In such cases, a less-finely tuned effective theory---namely, one that is less-finely tuned by virtue of the allowance of a greater range of parameter values that yield some value for an observable---is one that presents a larger target in its parameter space, into which some future higher-energy theory can `flow'. Of course, we contend that if we knew the higher-energy theory, it would provide a rationale (indeed, an explanation) for the particular parameter values that must be assumed for the effective theory to yield the correct value for the observable of interest. But when we do not know the higher-energy theory, it seems prudent to favor those effective theories that seem more flexible, in their capacity to describe observables of interest. Of course there is no guarantee that this is the right strategy, but it is one worthy of consideration. 

Does, then, our schema apply to `final theories'? There are two cases to consider here. First, if the theory has no free parameters then there is no fine-tuning and the depth of the explanation provided by the theory for any phenomenon (indeed, according to our measure for depth) is, as one might expect, unity. Secondly, if such a theory has free parameters  then it is not possible for our schema to assign a level of fine-tuning or depth. This is because it is difficult to justify the assignment of a range of possible values that could be taken by these parameters. This limitation of our approach reinforces our claims in the previous paragraph. Namely, restrictions on ranges of parameters that arise for effective theories relate to our expectations about higher-energy theories. For example, as described in~\cite{azhar+loeb_18}, the standard model of particle physics is not expected to apply at energy scales higher than the Planck energy scale, where we expect effects due to quantum gravity to be dominant. This expectation allows us to reasonably set an upper limit on the maximum energy scale for the standard model (which then translates into limits on parameters in the model). Now, if there is no expectation of the existence of a higher-energy theory---which is a defining feature of a final theory---and this final theory has free parameters, then it isn't clear that we can sensibly constrain the free parameters: our schema does not then provide a guide to computing levels of fine-tuning or of depth.

\subsection{A contrast with Hitchcock and Woodward~\cite{hitchcock+woodward_03}}\label{SEC:HWcf}

An interesting feature of our account of explanatory depth is its relationship to work by Woodward and Hitchcock~\cite{woodward+hitchcock_03} (on explanation) and Hitchcock and Woodward~\cite{hitchcock+woodward_03} (on explanatory depth). In short, their approach complements part of our construction while differing in other respects and it will be instructive to develop this comparison. Roughly, they define an explanation through a set of explanans variables, $\{X_1, X_2, \dots, X_n\}$, that enter into an explanatory generalization, $g$, which relates the explanandum variable, $Y$, to the explanans variables, where $Y=g(X_1, X_2, \dots, X_n)$. In particular, an explanation for the explanandum variable taking the specific value $y$ is an argument in which the explanans variables take some set of values, say $\{x_1, x_2, \dots, x_n\}$, such that when they enter into the explanatory generalization, one obtains the value $y$, namely, $y=g(x_1, x_2, \dots, x_n)$. A defining feature of explanatory generalizations is that they are ``invariant under testing interventions''~\cite[p.~17]{woodward+hitchcock_03}.  The invariance refers to a condition in which the explanatory generalization remains intact for counterfactual values of the explanans variables. The intervention is a process that brings about such a change in the values of the explanans variables (without directly affecting the explanandum variable), such as one that might be implemented by an experimenter (but an intervention does not have to invoke human agency). A testing intervention is an intervention that {\it does not} lead to the same value of the explanandum variable ($y$ in the example above). The depth of an explanation is tied to the ``range of invariance''~\cite[p.~182]{hitchcock+woodward_03} of the relevant explanatory generalization. A comparison of the depth of two explanations could advert to an assessment of, for example, the relative range of values of explanans variables over which their explanatory generalizations hold (or some comparison related to the topology of that range) or the accuracy of the two explanatory generalizations over their respective ranges of invariance (see~\cite{hitchcock+woodward_03} for a more complete account).

There is a clear sense in which there is a correspondence between our approach and the one just described. One can interpret our parameters, ${\bm p}$, as corresponding to the explanans variables while our observables, $\vec{O}$, correspond to (a higher-dimensional version of) the explanandum variable. A difference between these approaches arises in that we focus on particular types of relationships between parameters and observables (namely, what one may interpret as particular types of explanatory generalizations) and, moreover, in developing a notion of explanatory depth, we focus on a specific feature of such relationships. In particular, we focus on relationships such as those that one commonly finds for dynamical systems (see Sec.~\ref{SEC:Flatness}) or in certain phenomenological settings (see Sec.~\ref{SEC:Models}). We then analyze an aspect of depth that is not captured, we contend, by their subsequent work on depth (in~\cite{hitchcock+woodward_03}). That is, for any fixed number of parameters (and for a single observable, say), an important consideration (as regards depth) is the ranges of values of these parameters that do not yield significant changes in the value of the observable. In other words, the effective {\it invariance of the value of the explanandum variable itself}, for counterfactual values of the explanans variables, is a marker of a deep explanation. This sense of depth, we contend, closely follows the types of intuitions that are captured in theoretical developments in a broad range of physical settings.

\subsection{Further conceptual issues}\label{SEC:FurtherIssues}

Finally, let us highlight (and reiterate) some conceptual issues that are captured by our schema. Ceteris paribus, if the value of some explanandum is significantly affected by the addition of a single extra parameter to the explanans, the depth of the explanation decreases. It is only the case, according to our measure, that the depth of the explanation stays the same if the value of the explanandum is not significantly affected by values that the extra parameter can take. This is a desirable feature of our schema for it says that the depth of the explanation is not affected by the introduction of parameters in the explanans, with respect to which the explanandum is, in effect, independent.

There is a comparative role that can be played by our measure, in that two different explanations for common explananda can be compared. And when one can parse the observables of a model into, for example, distinct sets of observables, namely, distinct explananda, the depth of the explanations of these explananda can also be compared. In this way, our schema highlights a general functional role for explanatory depth, namely, it can be used in an operational setting to discriminate between models or to evaluate how well a single model explains different observables.\footnote{Note that in considering effective theories, our schema properly applies to the comparison of two such theories that are thought to apply at similar energy scales. We leave for future work the question of whether and how our approach could be adapted to compare the depth of two explanations, where one explanation involves a theory that reduces, in some limit, to the other theory. (See~\cite{weslake_10} for a related discussion of this general issue.)} [On this front, the examples we have selected in this paper refer to modern scientific models, but we expect our approach to work well in other cases from the history of science (see fn.~\ref{fn:examples}).]  A further operational role for our measure is as a catalyst for the search for new models. For example, if one was to find an upper bound on the depth of an explanation that was significantly less than unity, this would motivate the question of whether there is room for improvement using a different model.

It appears from our definition of the measure of depth [in Eq.~(\ref{EQN:Depth})], that a marker of a particularly deep explanation is one in which the observables are largely {\it independent} of the parameters, and so one might wonder why such parameters are to be included in our schema. On the other hand, it appears that the inclusion of parameters that are manifestly irrelevant to the system of interest (for example, the inclusion of parameters that track the prices of stocks, in a description of the evolution of spatial slices of the universe) does not affect the depth of the explanation; and so one might wonder why the inclusion of such parameters doesn't somehow decrease the depth of the explanation. Such objections ignore the context in which our account of explanatory depth has been developed. The parameters that we include play a salient role in the models in which they arise. For a dynamical system, the parameters are either initial conditions for dynamical variables that are necessary to describe the (physical) system of interest or else they are constants that are needed to make (physical) sense of the model, either internally or with respect to other models with which the model is supposed to cohere. For the types of phenomenological models we have explored above, each parameter can be thought of as tracking one of the empirical constraints that the system is deemed to satisfy.  Parameters that are not included are those not thought to contribute to our understanding of the system of interest as described by the model. We admit there are choices to be made in leaving out such parameters, but such choices are part of the practice of model building.\footnote{One goal that such model building (and scientific theorizing more broadly) seems attuned to is the pursuit of parsimonious descriptions of previously collected data, with a view to accurate future predictions. This was a major accomplishment of quantum chromodynamics, which explained a large amount of phenomenological data on nuclear physics (as mentioned in Sec.~\ref{SEC:ConnectingFTwithE}); and quantum electrodynamics, which achieved similar successes for electromagnetic phenomena. This goal also underlies the effort to find a theory that unifies all the forces of nature.}

More broadly, our schema for fine-tuning and explanatory depth can also be understood as one that relates `theoretical virtues' in the sciences (namely, a lack of fine-tuning and depth). Such a schema thus charts a middle ground between subjective elements (in that it refers to `virtues' or `values') and objective elements (in that such virtues are related to the construction of models that, at best, describe real features of the world). (See~\cite{mcmullin_82} for an elegant account of the tensions that arise in charting such a middle ground.) Unsurprisingly, there are context-dependent features of our account: one must choose a parameterization for a model and specify a range over which parameters may vary. Choices must also be made in determining which observables are salient. But having identified and fixed such choices, our schema endorses a simple quantitative way to connect fine-tuning and depth. It does so, we contend, in a way that is free from unnecessary complications and in a way that reflects the types of considerations that enter into the practice and development of the physical sciences.

\section*{Acknowledgements}
We thank Porter Williams for discussions. We acknowledge support from the Black Hole Initiative at Harvard University, which is funded through a grant from the John Templeton Foundation and the Gordon and Betty Moore Foundation. FA also acknowledges support from the Faculty Research Support Program (FY2019) at the University of Notre Dame.

\appendix
\section{Observables from probabilistic maps\label{SEC:AppendixA}}

The schema we have developed in Sec.~\ref{SEC:Explanation} can naturally be extended to the case where observables are probabilistically related to parameters (in a way that is distinct from our phenomenological account in Sec.~\ref{SEC:Models}). In particular, in the context of some physical model, $\mathcal{M}$, one can construct a probability density function for the observables given the parameters: $P(\vec{O}| \bm{p}, \mathcal{M})$. We assume also that we have access to a prior over parameters, namely, $P(\bm{p}|\mathcal{M})$. Instead of now restricting, to be finite, the domain over which $\bm{p}$ can take values, we assume, for the sake of simplicity, that $\bm{p}\in\mathbb{R}^{n}$ but that the prior, $P(\bm{p}|\mathcal{M})$, is nonzero only over a finite region. Similarly we assume that in principle, $\vec{O}\in\mathbb{R}^{m}$, but $P(\vec{O}| \bm{p}, \mathcal{M})$ is nonzero over some finite region. 

In this case therefore, an explanation of the vector of observables taking the value $\vec{O}_{M}$ (more specifically, taking a value in some small $m$-dimensional box, $\Delta_{\vec{O}_{M}}$, centered on $\vec{O}_{M}$) will comprise an argument in which the probability of the observables taking these values is larger than some fiducial value $\mu$.\footnote{Of course, we need to make precise how one determines this value. The variable $\mu$ is a stand-in for a `high value' and precisely what this value is (or should be) is a determination best made in the specific context in which an explanation is being constructed. We will not need to specify such a value to describe our scheme.} We can represent the resulting argument that comprises the explanation in the following summarized form:
\begin{equation}\label{EQN:EIS}
\bar{E}\;{\rm :}\;\bm{p}^{\prime} \wedge [P(\vec{O}_{M}| \bm{p}^{\prime}, \mathcal{M})\Delta_{\vec{O}_{M}} > \mu] \therefore \vec{O}_{M}.
\end{equation}

The depth of this explanation can be computed via the following steps.
\begin{itemize}
\item[(i)] For the $i$th parameter direction about the point $\bm{p}^{\prime}$, we construct the range, $\delta_i$, of parameter values over which observables do not change significantly. That is, the range over which the probability that the vector of observables takes values in the same small $m$-dimensional box described above, centered on $\vec{O}_{M}$, remains high: $P(\vec{O}_{M}| \bm{p}, \mathcal{M})\Delta_{\vec{O}_{M}} > \mu$.\footnote{Here we have assumed a small shift in the point in parameter space determined by a shift solely in the $i$th direction. We have represented the new point thus obtained by $\bm{p}$ instead of $\bm{p}^{\prime}$. By convention, limits of the range $\delta_i$ are determined by the {\it first} point in parameter space such that the probability of observables lying in $\Delta_{\vec{O}_{M}}$ is less than or equal to $\mu$.}
\item[(ii)] Next we find the probability of parameters lying in this range as gleaned from the appropriate marginal distribution:
\begin{equation}
P_{i}(\delta_i)\equiv \left[\prod_{k \neq i}\int_{\mathbb{R}}d{p_k}\right]\;\int_{\delta_{i}}dp_{i}\;P(\bm{p}| \mathcal{M}).
\end{equation}
\item[(iii)] Finally we define a corresponding measure of global fine-tuning, $\bar{\mathcal{G}}_{i}(\vec{O};\bm{p}^{\prime})$, extending the treatment in Sec.~\ref{SEC:Explanation} and in~\cite{azhar+loeb_18}: 
\begin{equation}
\bar{\mathcal{G}}_{i}(\vec{O};\bm{p}^{\prime})\equiv\log_{10}\left(\frac{1}{P_{i}(\delta_i)}\right).
\end{equation}
This measure is manifestly non-negative, with the minimum value (namely, zero) occurring when the probability of the observables lying in $\Delta_{\vec{O}_{M}}$, centered on $\vec{O}_{M}$, is greater than the fiducial value $\mu$, independent of the value of the $i$th parameter.
\end{itemize}

Our measure of the depth of the explanation in Eq.~(\ref{EQN:EIS}), which we denote by $\bar{D}_{\bar{E}}(\vec{O}; {\bm p}^{\prime})$ is then defined by the following:
\begin{equation}\label{EQN:Depth2}
\bar{D}_{\bar{E}}(\vec{O}; {\bm p}^{\prime})\equiv\frac{1}{\displaystyle \prod_{i=1}^{n} \left[1+\bar{\mathcal{G}}_{i}(\vec{O}; {\bm p}^{\prime})\right]}.
\end{equation}
This measure has analogous features to the measure described in Sec.~\ref{SEC:DepthMeas} [see points (i)--(v) under Eq.~(\ref{EQN:Depth})] with appropriate reassignments, for example, $\mathcal{G}\to\bar{\mathcal{G}}$ and $D_{E}\to\bar{D}_{\bar{E}}$.

\end{document}